\documentstyle[twocolumn,prl,aps,epsf]{revtex}
\begin{document}
\wideabs{ \draft
\title{Large magnetoresistance using hybrid spin filter devices}
\author{P. LeClair\footnotemark[1], J.K. Ha, H.J.M. Swagten, C.H. van de Vin, J.T. Kohlhepp, and W.J.M. de Jonge}
\address{Department~of~Applied~Physics and COBRA~Research~Institute, Eindhoven~University~of~Technology,
P.O.~Box~513, 5600~MB~~Eindhoven, The~Netherlands}
\maketitle
\begin{abstract}

A magnetic ``spin filter" tunnel barrier, sandwiched between a non-magnetic metal and a magnetic metal, is used to create a new magnetoresistive tunnel device, somewhat analogous to an optical polarizer-analyzer configuration. The resistance of these trilayer structures depends on the relative magnetization orientation of the spin filter and the ferromagnetic electrode. The spin filtering in this configuration yields a previously unobserved magnetoresistance effect, exceeding 100\%.

\end{abstract}

\pacs{PACS numbers: 73.40.GK, 75.70.-i, 85.30.Mn, 85.70.Kh}  }

\footnotetext[1]{Corresponding author. Electronic mail: pleclair@phys.tue.nl}


Spin electronic (``spintronic") devices~\cite{prinz}, based on utilizing the spin as well as the charge of electrons, open up an entirely new class of electronics. Such devices could include non-volatile magnetic memories, reprogrammable logic~\cite{prinz}, and quantum computers~\cite{vincenzo}. One thing hampering the development of spin electronic devices so far is the lack of sufficiently polarized (nearing 100\% spin polarization) current sources, for instance, for spin injection into semiconductors~\cite{molenkamp} or reprogramable logic~\cite{prinz}. So-called ``half-metallic ferromagnets," fully spin-polarized ferromagnets, would circumvent this problem~\cite{gustav_cro2}, but true half-metals have proven extremely difficult to realize in practice~\cite{moodera_hmf}.
However, rather than simply using a single nearly perfectly-polarized material, the phenomenon of spin filtering may also be exploited to create near 100\% polarization.
Here we propose and demonstrate a different approach, combining spin filter tunnel barriers~\cite{moodera88} and spin-dependent tunneling~\cite{spt,moodera95}, similar to a device proposed by Worledge {\it et al.}~\cite{double_filter}. The combination of a {\it non-magnetic} electrode with a spin filter tunnel barrier is used to effectively mimic a half-metallic tunneling electrode and achieve nearly 100\% spin polarization. Using this ``artificial half-metal" bilayer, we additionally use a second magnetic electrode, creating a nonmagnetic metal/ferromagnetic insulator/ferromagnetic metal (M-FI-F) device. We utilize EuS as the magnetic insulator, with Gd ferromagnetic and Al nonmagnetic electrodes. The tunnel current in this case depends on the relative magnetization orientation of the EuS filter and the Gd ``analyzer," in analogy to a half-metallic ferromagnet/insulator/ferromagnet tunnel junction. The spin filtering in this configuration yields a previously unobserved magnetoresistance effect, which we dub ``Spin Filter Injection Magnetoresistance" (SFIM), exceeding 100\%, suggesting a filtering efficiency close to 100\%. The present scheme would also circumvent impedance mismatch problems with semiconducting counter electrodes, and thus potentially allow spin injection from even a non-magnetic metal into a semiconductor. 

The principle of spin filtering with magnetic semiconductors has been demonstrated in field emission experiments~\cite{esaki}, and in tunnel junctions using a superconducting spin detector~\cite{moodera88}. The concept of spin filtering is illustrated in Fig.\ 1, using the well-known semiconducting Eu-chalcogenide, EuS~\cite{mauger,wachter}, which is ferromagnetic below T$_{\mbox{\scriptsize C}} \sim 16.8$K, as a spin filter. Above the T$_{\mbox{\scriptsize C}}$ of the EuS barrier, both spin-up and spin-down electrons experience the same potential barrier (Fig.\ 1a). Below T$_{\mbox{\scriptsize C}}$, due to the spin splitting of the conduction band in EuS (which forms the top of the tunnel barrier), the barrier height becomes spin-dependent, as shown in Fig.\ 1b. As a result of the exponential dependence of tunnel current on barrier height, one spin channel has a much larger tunneling probability than the other, resulting in a nearly 100\% spin-polarized current. 

With a magnetic metal, we must consider the role of the spin polarized density of states in the electrode as well. The tunnel current depends on the number of filled states in the first electrode as well as the number of available states in the second. Using one magnetic electrode, Fig.\ 1c, the density of available states in the magnetic electrode is spin dependent, and the tunnel current will depend on the relative orientation of the filtered spins ({\it i.e.}, the EuS magnetization direction) and the electrode magnetization. For parallel alignment, Fig.\ 1c, only majority (spin up) electrons tunnel through the filter, and thus they can only tunnel into majority states in the magnetic electrode, resulting in a large tunnel current. For the antiparallel case, (Fig.\ 1d), the current is minimal, since only the minority (spin down) states are available in the ferromagnet. One may consider this device analogous to a polarizer/analyzer optical configuration, albeit with a less-than-perfect analyzer, or a magnetic tunnel junction with one half-metallic electrode. The magnitude of the expected SFIM effect may be estimated within a simple two-current model~\cite{spt,simmons}, assuming spin conservation in the tunneling process, as $\Delta$R/R$_{\mbox{\scriptsize p}}$=2P$_{\mbox{\scriptsize m}}$P$_{\mbox{\scriptsize f}}$/(1-P$_{\mbox{\scriptsize m}}$P$_{\mbox{\scriptsize f}}$), where P$_{\mbox{\scriptsize m}}$ is the spin polarization of the ferromagnetic electrode, P$_{\mbox{\scriptsize f}}$ is the efficiency (polarization) of the spin filter, and $\Delta$R/R$_{\mbox{\scriptsize p}}$ is the change in resistance between parallel and antiparallel magnetization configurations normalized by the resistance in the parallel state. For a ferromagnetic electrode polarization of 50\%~\cite{spt}, and a filter efficiency of 90\%~\cite{moodera88}, one may expect a SFIM effect of more than 160\%. 

Devices were fabricated using conventional ultra-high vacuum sputtering techniques with in situ shadow masks onto oxidized Si(100) wafers. The EuS tunnel barrier was grown at 300$^\circ$C, while the metallic layers were grown at ambient temperature. Fig.\ 2 shows resistance versus magnetic field for a Si/SiO$_{2}$/Ta 5nm/Al 3nm/EuS 5nm/Gd 15nm structure at 2K (well below the EuS T$_{\mbox{\scriptsize C}}$), 30K (well above the EuS T$_{\mbox{\scriptsize C}}$), and at 7K (T/T$_{\mbox{\scriptsize C}}$$\sim$0.4). At 2K, an effect of $\sim$100\% (in some cases more than 130\%) is observed, clearly indicating the efficiency of the spin filtering. However, at 30K, above the EuS T$_{\mbox{\scriptsize C}}$, almost no magnetoresistance ($<$5\%) is observed, indicating clearly that the observed effects are due to the presence of a ferromagnetic spin filter barrier (the small persisting effect is due only to the field-induced magnetization in the EuS layer). Returning to the 2K data, for sufficiently high fields ($\sim$0.5T), when both magnetizations are parallel a low resistance state is observed. For some small fields, although the magnetization orientation in these structures is not completely well defined, a near antiparallel alignment is reached, and thus, a high resistance state is observed.  At 7K a large effect is still observed, and the switching behavior is also more controlled, though still a complete antiparallel alignment is not reached. For completely antiparallel alignment, an even larger SFIM effect is anticipated. Given the observed SFIM effects of more than 130\%, it can be determined using the aforementioned simple model that for a Gd polarization of 40\% or less~\cite{note.Gd}, our filter efficiency is essentially 100\%. Further, to indicate that a relative parallel or (nearly) antiparallel alignment is possible, we have measured magnetization versus magnetic field at 5K for separate EuS 5.0nm/Gd bilayers. Two distinct switching events are observed, one at low field corresponding to the EuS magnetization reversal, and one at higher fields corresponding to the Gd magnetization reversal. The presence of two distinct magnetization reversals indicates that the Gd electrode and EuS barrier can be switched independently (with some preliminary evidence for antiferromagnetic coupling between EuS and Gd), and may be aligned parallel or nearly antiparallel. 

Additional evidence of the nature of the spin filtering phenomenon can be obtained from the temperature dependence of the junction resistance. If spin filtering is present, one can expect the junction resistance to decrease as the temperature decreases~\cite{moodera88,esaki,mauger,wachter}. Specifically, below T$_{\mbox{\scriptsize C}}$ the barrier height for spin up electrons is much lower (by $\sim$0.18eV~\cite{mauger,wachter}) than for either spin up or spin down electrons above T$_{\mbox{\scriptsize C}}$. Thus, spin up electrons preferentially tunnel because of this lowered barrier, leading to a resistance decrease at low temperatures in addition to the high spin polarization. More quantitatively, the tunnel resistance (for a vanishing external bias) can be expressed within a simple free-electron tunneling model~\cite{moodera88,spt,simmons} as:

\begin{equation}
R_{\uparrow\downarrow} \sim \exp\left( d \; \varphi_{\uparrow\downarrow}^{\frac{1}{2}}(T)\right) \;\;\;\; \varphi_{\uparrow\downarrow}(T) = \overline{\varphi} \mp J_{df} S \sigma(T)
\end{equation}

where $d$ is the barrier thickness, T is the temperature, $\overline{\varphi}$ is the average tunnel barrier height, $J_{df}$ is the d-f exchange constant for EuS~\cite{mauger}, S=7/2 is the spin quantum number of a Eu$^{2+}$ ion , and $\sigma(T)$ is the reduced magnetization M(T)/M(T=0) of EuS, with the $\uparrow (\downarrow)$ are to denote spin up(down) electrons. We expect, then, that the temperature dependence of the tunnel resistance should scale (exponentially) with the magnetization of the EuS filter. Shown in Fig.\ 3 is the normalized tunnel resistance as a function of temperature, in several applied fields, for an Al/EuS/Gd junction. Indeed, a clear decrease of the tunnel resistance is observed below the T$_C$ of EuS, which provides proof that spin filtering is present. Further, the broadening of this resistance transition with increasing applied field is expected from the smearing of the magnetization-temperature behavior near the transition in the presence of an external field~\cite{ohandley}. This clearly indicates that the resistance transition is related to the magnetic phase transition in EuS. Further, the exponential sensitivity of the tunnel resistance to the reduced magnetization of the EuS layer may explain the noise observed in the R(H) data, as small fluctuations in magnetization are amplified in the resistance signal. We note parenthetically that although the absolute resistance of these devices is quite high, as expected for a 50\AA\ barrier, reducing the EuS thickness by a factor of two would gain several orders of magnitude in resistance, without seriously affecting the magnetic quality~\cite{thin_eus}. 

Finally, to indicate that all of the observed phenomena are genuinely due to tunnel transport, in the inset of Fig.\ 3 we show a conductance-voltage (dI/dV-V) curve at 5K for an Al/EuS/Gd junction. A roughly parabolic and symmetric behavior is observed at higher voltages, indicative of tunnel transport~\cite{simmons}. At low voltages, a linear contribution is observed, consistent with tunneling assisted by magnon excitations in the Gd layer~\cite{zhang.bd}. Fitting the I-V curves to Simmons~\cite{simmons} model gives barrier heights of $\sim$0.5eV below the EuS T$_{\mbox{\scriptsize c}}$. By fitting curves below and above T$_{\mbox{\scriptsize c}}$, we obtain an exchange splitting of $\approx$0.36eV (Using Eq. 1), in agreement with bulk data~\cite{mauger,wachter} and previous tunneling measurements~\cite{moodera88}. Though the barrier heights are much lower than previously reported values~\cite{moodera88}, this is probably a result of the defective and polycrystalline nature of our sputtered EuS films. Additionally, at temperatures above the EuS T$_{\mbox{\scriptsize C}}$, the tunnel resistance decreases as temperature increases, consistent with of tunnel transport. We may conclude that we are indeed observing magnetoresistance due to spin filtering tunneling via the EuS tunnel barrier. 

Summarizing, we report a large and newly discovered magnetoresistance effect in Al/EuS/Gd structures, resulting from the combination of a spin filter tunnel barrier and a ferromagnetic electrode. These ideas may have potential utility for spin injection into semiconductors or for novel hybrid devices. The large polarization achievable using spin filters~\cite{moodera88}, as well as the lack of any impedance mismatch problem~\cite{molenkamp} with semiconductors, makes spin filtering a nearly ideal method for spin injection into semiconductors, enabling novel spintronic devices~\cite{prinz}. Finally, a spin-filter operating at room temperature would be possible, utilizing the myriad of ferro- and ferri- magnetic ferrites and garnets~\cite{ohandley}, or potentially the recently predicted room temperature diluted magnetic semiconductors~\cite{dietl}, which can be made magnetic and insulating above room temperature.

The authors would like to acknowledge J.S. Moodera for helpful discussions. P. LeClair and J.K. Ha are supported by the Netherlands technology foundation STW, and we also acknowledge support from the Dutch foundation for condensed matter research FOM.

\begin{figure}[htb]
\begin{center}
\hspace{0cm}\epsfxsize=0.95\columnwidth \epsfbox{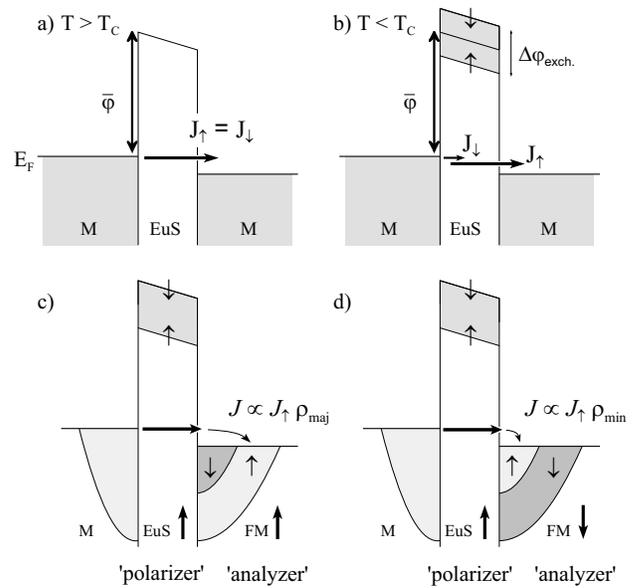}
\caption[]{Schematic illustration of spin filtering and the magnetoresistance effect. (a), above T$_{\mbox{\scriptsize C}}$ of the EuS filter the two spin currents are equal. (b), below the T$_{\mbox{\scriptsize C}}$ of EuS, the tunnel barrier is spin-split, resulting in a highly spin polarized tunnel current. With a ferromagnetic (FM) electrode, the tunnel current depends on the relative magnetization orientation. For parallel alignment (P), (c),  a large current results, while for antiparallel alignment (AP), (d), a small current results.
 } \label{fig1}
\end{center}
\end{figure}

\begin{figure}[htb]
\begin{center}
\hspace{0cm}\epsfxsize=0.95\columnwidth \epsfbox{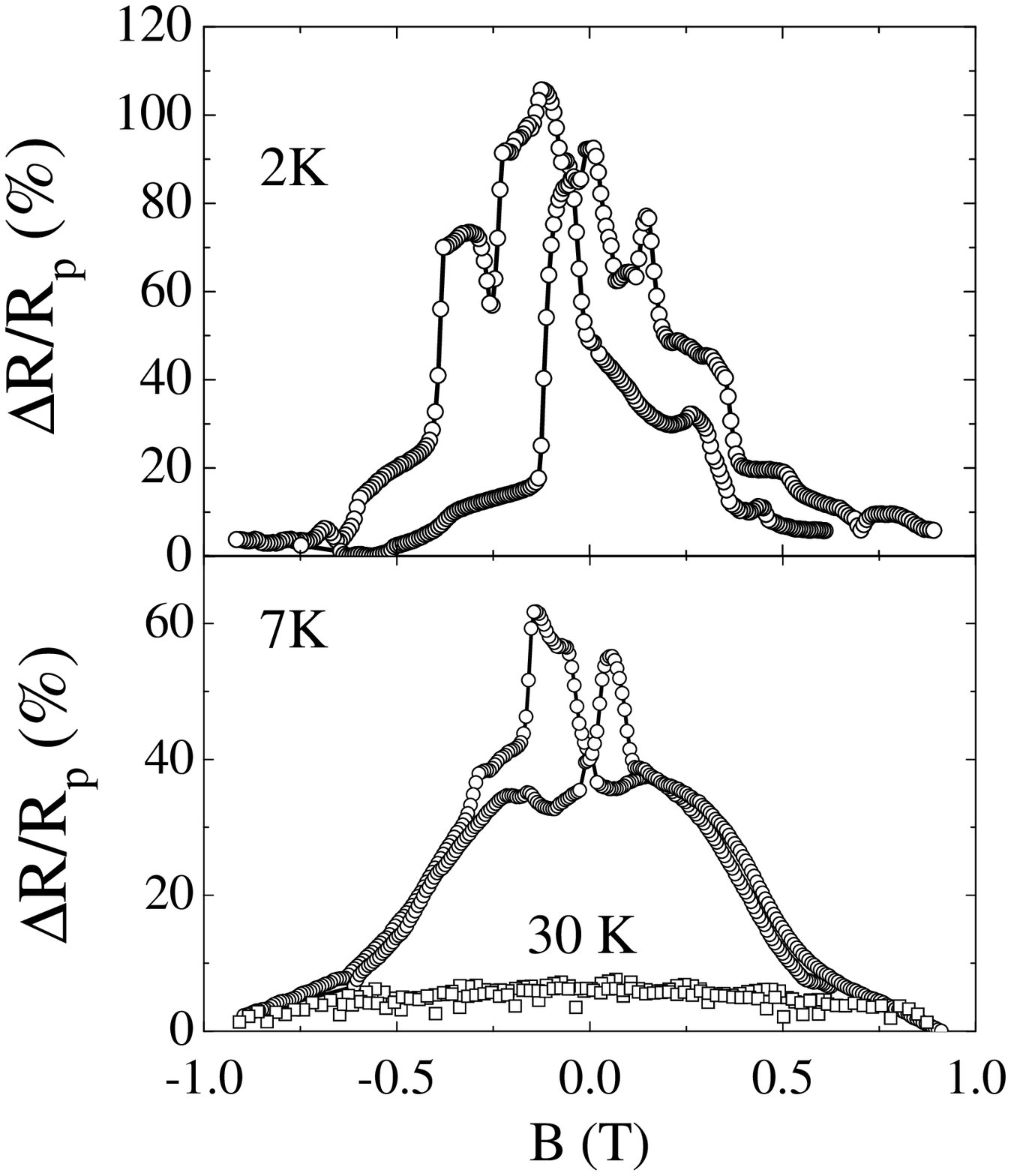}
\caption[]{(a) Magnetoresistance as a function of magnetic field at 2K (well below the EuS T$_{\mbox{\scriptsize C}}$), (b) at 7K, and at 30K (well above the EuS T$_{\mbox{\scriptsize C}}$).} \label{fig2}
\end{center}
\end{figure}

\begin{figure}[htb]
\begin{center}
\hspace{0cm}\epsfxsize=0.95\columnwidth \epsfbox{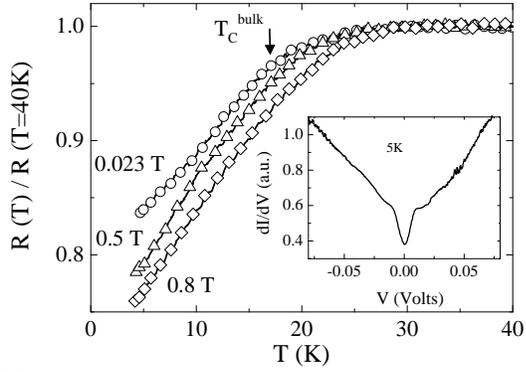}
\caption[]{Normalized resistance {\it vs.} temperature behavior for several values of magnetic field: B=0.0T (circles), B=0.3T (squares), and B=0.8T (triangles). Inset: conductance-voltage (dI/dV-V) characteristics at 5K for a representative Al/EuS/Gd device.} \label{fig3}
\end{center}
\end{figure}


\begin{thebibliography}{10}
\bibitem{note.Gd}
The Gd tunneling spin polarization is yet unknown with EuS barriers. We believe
  that the polarization of 14\% for Al/Al$_2$O$_3$/Gd quoted in Ref.\ [7] would
  increase considerably if repeated under UHV conditions (as reported for Ni,
  Ref.\ [7]).
\end{thebibliography}
\end{document}